\documentclass[sigconf]{acmart}

\AtBeginDocument{%
  }

\usepackage[utf8]{inputenc}

\usepackage[para,online,flushleft]{threeparttable}
\usepackage[english]{babel}
\usepackage{amsthm}
\usepackage{amsmath}
\usepackage{varwidth}
\usepackage{subcaption}
\usepackage{calc}
\usepackage{blindtext}
\usepackage{multicol}
\usepackage{graphicx}
\usepackage{paralist}
\usepackage[english]{babel}
\usepackage{float}
\usepackage{makecell}
\usepackage{multirow}
\usepackage{amsfonts}
\usepackage{booktabs}
\usepackage{siunitx}
\usepackage{etoolbox}
\usepackage{numprint}
\usepackage[T1]{fontenc}
\usepackage{algorithm}
\usepackage{flushend}
\usepackage[noend]{algpseudocode}
\usepackage{textcomp}
\usepackage{xcolor}
\usepackage{fancyvrb}
\usepackage{paralist}
\usepackage[inline]{enumitem}
\usepackage{booktabs}
\usepackage{color}
\usepackage{colortbl}
\usepackage{flexisym}
\usepackage{array}
\usepackage{xparse}
\usepackage{xspace}
\usepackage[flushleft]{threeparttable}
\usepackage[nobreak]{mdframed}
\usepackage{framed}
\usepackage{pifont}
\usepackage{balance}

\newcommand{\eg}{\hbox{\emph{e.g.}}\xspace}

\usepackage{listings}
\usepackage[]{xcolor}

\theoremstyle{definition}

\newcommand{\toolname}{\textsc{CppPerf-Mine}\xspace}
\newcommand{\db}{\textsc{CppPerf-DB}\xspace}
\newcommand{\openhands}{\textsc{OpenHands}\xspace}
\newcommand{\mycomment}[1]{}

\usepackage[]{xcolor}
\newcommand{\TODO}[1]{\textcolor{red}{#1}\GenericWarning{}{LaTeX Warning: TODO: #1}}\newcommand\todo\TODO

\newcommand{\MODIFIED}[1]{\textcolor{blue}{#1}}\newcommand\modified\MODIFIED

\algnewcommand{\Inputs}[1]{%
  \Statex \textbf{Inputs:}
  \Statex \hspace*{\algorithmicindent}\parbox[t]{.8\linewidth}{\raggedright #1}
}
\algnewcommand{\Outputs}[1]{%
  \Statex \textbf{Outputs:}
  \Statex \hspace*{\algorithmicindent}\parbox[t]{.8\linewidth}{\raggedright #1}
}
\algnewcommand{\Initialize}[1]{%
  \State \textbf{Initialize:}
  \Statex \hspace*{\algorithmicindent}\parbox[t]{.8\linewidth}{\raggedright #1}
}

\definecolor{javared}{rgb}{0.6,0,0} 
\definecolor{javagreen}{rgb}{0.25,0.5,0.35} 
\definecolor{javapurple}{rgb}{0.5,0,0.35} 
\definecolor{javadocblue}{rgb}{0.25,0.35,0.75} 

\lstdefinestyle{diff}{
    escapechar=\%
}

\usepackage{hyperref}
\addto\extrasenglish{%

}

\usepackage{cleveref}

\setcopyright{none}
\renewcommand\footnotetextcopyrightpermission[1]{} 
\begin{document}

\title{CppPerf: An Automated Pipeline and Dataset for Performance-Improving C++ Commits}

\author{Tommy Ho}
\email{tho@student.ethz.ch}
\affiliation{
  \institution{ETH Zurich}
  \country{Switzerland}
}

\author{Khashayar Etemadi}
\email{ketemadi@ethz.ch}
\affiliation{
  \institution{ETH Zurich}
  \country{Switzerland}
}

\author{Zhendong Su}
\email{zhendong.su@inf.ethz.ch}
\affiliation{
  \institution{ETH Zurich}
  \country{Switzerland}
}


\begin{abstract}
Recent progress in automated repair of performance bugs demands realistic, executable benchmarks. However, existing C++ performance benchmarks are largely built from competitive programming submissions, and recent real-world benchmarks predominantly target Python and .NET. To fill this gap, we present \toolname, a configurable pipeline that mines execution-time-improving patches from open-source C++ repositories on GitHub by combining structural commit filtering, an LLM-based commit classifier, and a containerized build \& test stage that produces fully reproducible Docker images for each patch. Using \toolname, we build \db, a benchmark comprising 347 manually verified patches from 42 mature C++ repositories, 39\% of which are multi-file, enabling the evaluation of repository-level repair tools. In our preliminary study, \openhands correctly fixes only 13.5\% of the patches in \db, confirming that real-world C++ performance repair remains an open challenge.

\toolname and \db are open-source and publicly available at: \url{https://doi.org/10.5281/zenodo.20097425}.
 
In addition, a demonstration video is available at:\\ \url{https://www.youtube.com/watch?v=nixlupIgSdM}.
\end{abstract}


\maketitle

\section{Introduction}
\label{sec:intro}

After years of advancements in the field of automated program repair (APR), which has been mainly focused on fixing functional bugs~\cite{just2014defects4j,Jimenez2023SWEbenchCL}, recent research has started to build new techniques for automated repair of performance bugs~\cite{garg2025rapgen,ren2025peace,gao2025search}.
One of the challenges in this area is the scarcity of benchmarks of real-world performance bug-fixing patches. The benchmarks of performance patches that are most commonly used for evaluating APR tools contain online competition programs~\cite{Madaan2023LearningPC,huang2024effibench,du2024mercury}, introductory-level programs~\cite{liu2024evaluating}, and generated programs~\cite{huang2024effilearner}. Recently, several benchmarks have been proposed that are collected from real-world projects in Python~\cite{he2025swe, ma2025swe, shetty2025gso, ren2025peace}, C\#~\cite{garg2025perfbench}, Java~\cite{yi2025experimental}, and C++~\cite{Madaan2023LearningPC}. However, they are typically collected from a small number of projects and provide no configurable tool for extending them according to user needs~\cite{ma2025swe, yi2025experimental}.

In this paper, we introduce \toolname, a novel and configurable tool for building benchmarks of executable performance bug-fixing patches in C++ repositories. \toolname searches the commit history of GitHub C++ repositories, filters them based on user-provided configurations, employs an LLM-based technique to identify commits that improve execution time, checks that the project builds and passes tests, and outputs the executable execution-time-improving commits as Docker images that provide full reproducibility. With this approach, \toolname can build extensible benchmarks by collecting additional and newer commits.

Using \toolname, we build \db, a benchmark with 347 real-world executable C++ patches that fix execution-time bugs and are collected from 42 different repositories. The patches in \db are collected from repositories that have between 306 and 28,718 stars, and 39\% (136/347) of them modify multiple files, making \db diverse and suitable for evaluating repository-level patch generation. Our preliminary study shows that \openhands, a state-of-the-art agentic tool, fixes only 13.5\% of the bugs in \db, indicating the challenging nature of automated repair of real-world performance bugs.

In summary, we make the following contributions:
\begin{itemize}
\item \textbf{\toolname}, a configurable tool for building benchmarks of executable C++ performance bug-fixing patches.

\item \textbf{\db}, a benchmark with 347 real-world C++ execution time bug-fixing patches collected from 42 repositories that span 6 years of development history.

\item An \textbf{evaluation} of the diversity of patches in \db and the precision of \toolname, and a preliminary study using \db to assess the effectiveness of state-of-the-art agentic tools in fixing C++ performance bugs.
\end{itemize}

\section{\toolname: The Pipeline}
\label{sec:pipeline}
\toolname takes user-defined settings as input and uses them to find commits that fix execution time bugs in C++ repositories. For each such commit, it produces a containerized environment that builds and runs tests on the corresponding patch.

\begin{figure*}
\begin{center}
\includegraphics[width=\textwidth]{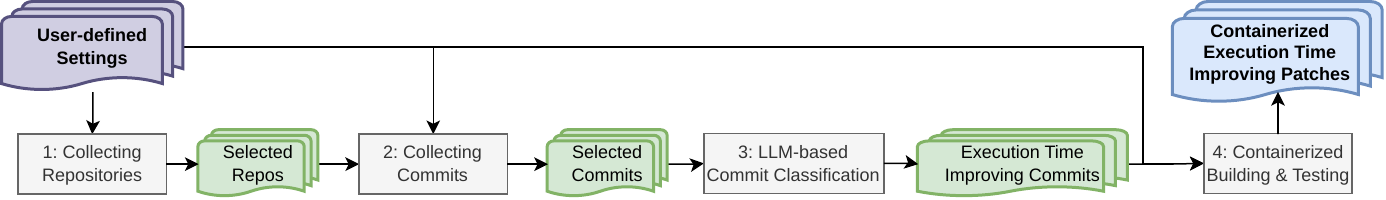}
\caption{An overview of the \toolname workflow.}
\label{fig:cppperfmine}
\end{center}
\end{figure*}

\autoref{fig:cppperfmine} shows an overview of the \toolname workflow, which has four main components: (1) collecting repositories, (2) collecting commits, (3) LLM-based commit classification, and (4) containerized building and testing. All components consider user-defined settings to select and build containerized execution-time-improving patches that meet user expectations.

\subsection{Collecting Repositories and Commits}
\label{sec:collection}
\toolname starts by crawling GitHub to select repositories that meet five requirements. First, the repository should have at least the minimum number of stars specified by the user to ensure that it is a solid and popular project. Second, C++ should be the primary language of the repository. Third, the project should have a \texttt{CMakeLists.txt} file at its root and use CMake to build, allowing us to build and test selected projects with a unified technique. Fourth, the repository must define executable tests within its CMake configuration, which are used to check that the patch does not break the functionality of the program. Fifth, we run the tests on the latest version of the project to confirm that they build and pass successfully.

After selecting the repositories that meet the requirements, \toolname examines their commit histories and filters commits according to three \textit{structural criteria}. First, the commit should be from the user-defined period of time. Second, the commit should not modify more than the maximum number of files specified by the user. Finally, the commit should only change C++ source code files; \toolname excludes commits that change test files to ensure that a fixed set of tests can serve as a specification of program functionality in both the original and patched versions. Commits that meet these criteria are selected and passed on to the next step.

\subsection{LLM-based Commit Classification}
\label{sec:llm-based-classification}

\toolname uses a two-phase LLM-based technique to assess whether a selected commit focuses on improving execution time. In the first phase, two LLMs are presented with the commit message and, if available, the description of the linked issue, and asked to determine if the commit is focused on execution time improvement by answering ``Yes'', ``No'', or ``Maybe''. If both LLMs agree on ``Yes'' or ``No'', we take that label as the final classification. This ensures that we classify a commit based on its commit message and linked issue description only when two LLMs are highly confident.

If the LLMs do not agree, we proceed to the second phase, in which a single LLM is presented with the commit message, the issue description, and the code diff, and asked to respond with either ``Yes'' or ``No''. The code diff provides detailed information about the modifications beyond the high-level commit message and issue description, but it can be large and costly to process. Therefore, we include it only for the limited number of commits that reach the second phase.

\subsection{Containerized Building \& Testing}
\label{sec:build-test}

In the final step, \toolname builds and runs the tests for each commit classified as execution-time-improving in a dockerized environment. It first fetches a base Docker image with an appropriate version of GCC and clones the original and patched versions of the repository into separate directories. Then it applies various heuristics to install the libraries required to build the project, falling back to an LLM-based iterative technique to detect and install missing libraries if the build fails. Finally, \toolname builds and runs the tests on both versions a user-specified number of times to ensure that the build \& test process is consistently successful. If both versions build and pass all runs, the result is saved as a Docker image containing the required compiler, libraries, both versions of the code, and execution logs. This image is the output of \toolname as a ``containerized execution time improving patch''.

\subsection{Implementation}
\label{sec:implementation}
\toolname is programmed in Python. By default, it collects commits from 2020 to 2025 in repositories with at least 300 stars and changing at most 20 files. The LLM-based commit classifier uses \texttt{qwen2.5:7b} and \texttt{qwen3:8b}, with the latter also used in the second phase. These open-source models are chosen to enable local deployment. In the containerized building and testing step, the tests are executed 31 times by default; the first execution is treated as a warm-up and discarded. Following the previous work~\cite{he2025swe}, a test is considered to demonstrate a statistically significant execution time improvement if it shows at least a 5\% improvement with a p-value below 0.05 in Mann-Whitney test. \toolname and \db are publicly available on both Zenodo and GitHub~\cite{repo}.

\subsection{Usage}
\label{sec:usage}
Both \toolname and \db are used through a command-line interface provided by a \texttt{main.py} script. To build a benchmark, users supply their desired filters and configurations as arguments, and \toolname produces a set of Docker images together with a JSON file per image describing the patch source and execution data. To evaluate a patch generation tool against \db, users pull the Docker image of a selected patch and invoke \texttt{main.py} with the patch id and their generated patch; the script runs the tests and saves the results to a JSON file. Collected patches and evaluation results can be inspected via the same CLI.

\section{Experiments}
\label{sec:experiments}

To evaluate \toolname, we begin by running it with its default configuration to build \db, an extensible benchmark of real-world execution time bug-fixing patches in C++ projects. Using \toolname and the resulting \db, we then answer three research questions:

\newcommand\rqone{What are the characteristics of the patches in \db in terms of scale, diversity, and scope?}
\newcommand\rqtwo{To what extent can off-the-shelf advanced coding agents fix execution time bugs in \db?}
\newcommand\rqthree{How precise is \toolname for building benchmarks of C++ execution time bug-fixing patches?}

\textbf{RQ1} (\db characteristics): \rqone

\textbf{RQ2} (\db in practice): \rqtwo

\textbf{RQ3} (\toolname precision): \rqthree

\subsection{RQ1 Experiment (\db characteristics)}
\label{sec:rq1}

In total, \toolname scans 65,942 commits, of which 25,715 meet the structural criteria and are passed to the LLM-based classifier. The classifier identifies 1,120 commits as focused on execution time bugs, and \toolname successfully builds and runs tests for 493 of them in a fully reproducible containerized environment. We then manually verify these executable patches: a patch is retained if (a) the commit message or linked issue explicitly mentions speed improvement, or (b) the code diff contains a recognizable optimization pattern (e.g., algorithmic improvement, caching, reduced allocations). This step excludes 146 patches that are not clearly focused on execution time improvement and might add noise to the evaluation of patch generation tools. Consequently, \db consists of 347 manually verified, real-world execution time bug-fixing patches that build and pass all tests in a reproducible environment. For 35 of these patches, an existing test case demonstrates a significant improvement in execution time and can be used for automated and objective assessment of generated patches.

\begin{table}[t]
\centering
\begin{tabular}{lrrrrr}
\hline
\textbf{Metric} & \textbf{Min} & \textbf{Q1} & \textbf{Median} & \textbf{Q3} & \textbf{Max} \\
\hline
Commits & 126 & 2,032 & 3,468 & 5,829 & 14,428 \\
Stars & 306 & 1,307 & 1,849 & 5,493 & 28,718 \\
\hline
\end{tabular}
\caption{The number of commits and stars across the repositories from which the 347 patches in \db are collected.}
\label{tab:commit_star_summary}
\end{table}

The 347 patches in \db are collected from 42 repositories. As shown in \autoref{tab:commit_star_summary}, these repositories have between 126 and 14,428 commits (median 3,468) and between 306 and 28,718 stars (median 1,849), indicating a diverse mix of well-established and mature projects. This places \db well beyond the toy programs considered in widely-used benchmarks~\cite{Madaan2023LearningPC}. Furthermore, 39\% (136/347) of the patches are multi-file, making \db suitable for evaluating repository-level patch generation tools~\cite{wang2025openhands,yang2024swe,zhang2024autocoderover}, and patches modify between 1 and 6,069 lines of code (median 28), spanning a wide range of scopes.

\begin{mdframed}\noindent
    \textbf{Answer to RQ1:} \\
    \db contains 347 containerized execution time improvement patches from 42 repositories with a median of 1,849 stars and 3,468 commits, indicating their maturity and popularity. 39\% of the patches modify multiple files, enabling the evaluation of repository-level patch generation tools.
\end{mdframed}

\subsection{RQ2 Experiment (\db in practice)}
\label{sec:rq2}

\begin{table}[t]
\centering
\caption{\openhands results on \db.}
\label{tab:openhands-results}
\begin{tabular}{lrrr}
\toprule
\textbf{Result} & \makecell{\textbf{All}\\(n=347)} & \makecell{\textbf{Single-file}\\(n=211)} & \makecell{\textbf{Multi-file}\\(n=136)} \\
\midrule
Correct Patch   & 47 (13.5\%)   & 37 (17.5\%)   & 10 (7.4\%)   \\
Correct Location             & 101 (29.1\%)  & 57 (27.0\%)   & 44 (32.4\%)  \\
Incorrect & 199 (57.3\%)  & 117 (55.5\%)  & 82 (60.3\%)  \\
\bottomrule
\end{tabular}
\end{table}

We perform a preliminary study to assess the effectiveness of \openhands, a state-of-the-art patch generation tool, on \db. For each patch, we give \openhands the original code, the commit message, the issue description (if available), and the list of modified files of the ground-truth patch, and ask it to generate a patch that improves execution time. We provide the list of modified files to constrain the search space and reflect realistic development scenarios, and configure \openhands to use \texttt{gpt-5-mini}. We then manually compare each generated patch with its ground-truth: a patch is \textit{correct patch} if it performs the same semantic modification, \textit{correct location} if it modifies the same code but with different semantics, and \textit{incorrect} otherwise.

\autoref{tab:openhands-results} shows the results. \openhands generates a semantically equivalent patch for 13.5\% (47/347) of the bugs in \db, including 17.5\% of single-file and 7.4\% of multi-file bugs. This indicates that \db contains challenging bugs, especially among its multi-file ones, that are suitable for evaluating tools aimed at advancing the state of the art. The table also shows that in 57.3\% of the cases, the generated patch does not even correctly locate the bug, reaffirming the importance of accurate fault localization in fixing performance bugs.

\begin{mdframed}\noindent
    \textbf{Answer to RQ2:} \\
    \openhands generates a patch semantically matching the ground-truth for 13.5\% (47/347) of performance bugs in \db, indicating that \db contains many challenging bugs suitable for evaluating advanced patch generation tools.
\end{mdframed}

\subsection{RQ3 Experiment (\toolname precision)}
\label{sec:rq3}

\begin{table}[t]
\centering
\caption{Confusion matrix of \toolname's LLM-based classifier on a manually labeled sample of 405 commits. It indicates a precision of 86.67\%.}
\label{tab:classifier-confusion}
\begin{tabular}{lccc}
\toprule
\textbf{Manual Label} & \makecell{\textbf{Predicted} \\ \textbf{Positive}} & \makecell{\textbf{Predicted} \\ \textbf{Negative}} & \textbf{Total} \\
\midrule
Positive & 13 (TP)  & 18 (FN)  & 31  \\
Negative & 2 (FP)   & 372 (TN) & 374 \\
\midrule
Total    & 15       & 390      & 405 \\
\bottomrule
\end{tabular}
\end{table}

To assess whether the LLM-based classifier of \toolname correctly labels commits, we randomly select and manually analyze 405 commits passed to the classifier. \autoref{tab:classifier-confusion} shows the resulting confusion matrix. \toolname predicts 15 commits as positive, 13 of which are manually verified to focus on improving execution time, yielding a precision of 86.67\%. The classifier's recall is 41.94\%, showing that a notable number of execution time improvement patches are discarded. While higher recall is desirable, we consider precision more important for a patch collection tool: a patch included in the benchmark should be highly likely to be useful for evaluation. We therefore intentionally design a conservative classifier that favors precision, even at the cost of excluding some execution-time-improving patches.

\begin{mdframed}\noindent
    \textbf{Answer to RQ3:} \\
    Manual analysis of 405 randomly selected commits shows that the LLM-based classifier of \toolname has a precision of 86.67\%, making it a reliable tool for finding execution-time-improvement patches.
\end{mdframed}

\section{Threats to Validity}

\textit{External Validity:} \toolname considers C++ projects that use CMake and relies on heuristics and LLM-based techniques to build them, excluding projects that use other build tools or that cannot be built with our approach. Extending \toolname to handle additional build systems and projects dismissed by the current version is an opportunity for future work.

\vspace*{3pt}
\noindent
\textit{Construct Validity:} \toolname relies on test execution times across multiple runs to decide if a test demonstrates an improvement. As reported in \autoref{sec:rq1}, only 35 of the 347 patches in \db have an existing test that detects execution time improvement, reflecting a known limitation: human-labeled performance patches may target code paths not exercised by existing tests, or improvements may be too small to surface above measurement noise. To mitigate this, \toolname employs the configurable statistical testing described in \autoref{sec:build-test}, which increases confidence in the tests that demonstrate improvement.

\section{Related Work}
\label{sec:related-work}

\textbf{Performance Benchmarks.}
Traditional APR benchmarks such as Defects4J~\cite{just2014defects4j} and SWE-Bench~\cite{Jimenez2023SWEbenchCL} target functional defects rather than performance issues. Early performance benchmarks are built from competitive programming submissions, including PIE~\cite{Madaan2023LearningPC}, EffiBench~\cite{huang2024effibench}, and Mercury~\cite{du2024mercury}, but consist of small, self-contained programs that differ substantially from real-world software~\cite{coignion2024performance}. More recent benchmarks mine performance patches from real repositories: SWE-Perf~\cite{he2025swe}, SWE-fficiency~\cite{ma2025swe}, GSO~\cite{shetty2025gso}, PeacExec~\cite{ren2025peace}, and PerfBench~\cite{garg2025perfbench}. Despite C++ being a dominant language for performance-critical software, the only widely-used C++ performance benchmark remains PIE~\cite{Madaan2023LearningPC}, built from online competition submissions. \db addresses this gap with 347 manually verified, containerized patches mined from 42 real-world C++ repositories, and \toolname enables extending the benchmark with newly mined commits.

\vspace*{3pt}
\noindent
\textbf{Automated Performance Repair.}
Function-level methods include RAPGen~\cite{garg2025rapgen}, SBLLM~\cite{gao2025search}, and EFFI-EARNER~\cite{huang2024effilearner}, while PEACE~\cite{ren2025peace} targets project-level optimization. General-purpose agentic frameworks, e.g., \openhands~\cite{wang2025openhands}, SWE-Agent~\cite{yang2024swe}, AutoCodeRover~\cite{zhang2024autocoderover}, have also been evaluated on performance tasks, with reported success rates remaining low (\eg, 4.9\% on GSO and 3\% on PerfBench)~\cite{shetty2025gso,garg2025perfbench}. Our preliminary study of \openhands on \db reports 13.5\%, consistent with these results and reaffirming that real-world C++ performance repair remains an open challenge.

\section{Conclusion}
\label{sec:conclusion}
 
We presented \toolname, a configurable pipeline that mines reproducible execution time improving patches from open-source C++ repositories, and \db, a benchmark of 347 such patches spanning 42 mature projects, 39\% of which modify multiple files. A preliminary evaluation shows that \openhands semantically matches the ground-truth for only 13.5\% of bugs in \db, confirming that real-world C++ performance repair remains an open challenge. We release \toolname so that the community can extend \db with newer commits and tailor benchmarks to their evaluation needs.

\bibliographystyle{ACM-Reference-Format}
\balance
\bibliography{references}

\end{document}